\documentstyle[12pt]{article}
\begin{document}
\thispagestyle{empty}
\begin{center}
\LARGE \tt \bf {Torsion effects on vortex filaments and Hasimoto soliton transformation in magnetars}
\end{center}
\vspace{3.5cm}
\begin{center}
{\large By L.C. Garcia de Andrade\footnote{Departamento de F\'{\i}sica Te\'{o}rica - IF - UERJ - Rua S\~{a}o Francisco Xavier 524, Rio de Janeiro, RJ, Maracan\~{a}, CEP:20550.e-mail:garcia@dft.if.uerj.br}}
\end{center}
\begin{abstract}
The role played by torsion of magnetic vortex line curves or filaments, in the equilibrium state of magnetars is investigated. When the magnetars equilibrium equations are written in terms Frenet-Serret frame it is shown that in regions of the magnetic star where the Frenet torsion is constant it induces an oscillation in the vortex filaments. By solving the magnetar equilibrium equation we shown the this behaviour also appears in the magnetic field. The first derivative of the gravitational potential with respect to the arc lenght of the vortex filament is shown to coincide with the Hasimoto soliton transformation of the Schroedinger equation for the constant torsion.
\end{abstract}
\newpage
\section{Introduction}
The role of torsion effects in twisted vortex filaments have been recently investigated in detail by Ricca \cite{1}. In his work he demonstrated that torsion influences cconsiderably the motion of helical vortex filaments in an imcompressible perfect fluid, where the binormal component is responsible for the displacement of the vortex filament in the fluid. Ricca's prescription follows the Moore-Saffman \cite{2} to helices of any pitch. More recently \cite{3} Ricca has discussed the problem of inflexional desequilibrium of inflexional magnetic-flux tubes, mechanism which has important implications energy balance in solar coronal loops \cite{4} and astrophysical flows. In this work we solve the equilibrium equations of a magnetic stars ,or in brief magnetars, by addopting the method of vortex filaments and writing the magnetar equations in the Serret-Frenet frame. In this approach we are able to investigate the vortex filaments inside the magnetars by making use of the constant torsion approximation. Vortex filaments are known to exists in the interior of the Sun which is a classical example of magnetars \cite{5}. Therefore by considering regions of magnetars where torsion is constant, we are able to show that the magnetic field oscillates in the helical form. This region can give rise to turbulent flows. The paper is organised as follows: in section II we present the model of the magnetars. In section III we present the solution of the equation based on the vortex magnetic lines where the currents and the magnetic field appear in terms of the torsion of the vortex filaments of the star. Section IV presents the discussions.        
\section{Magnetars in Equilibrium}
The idea of the magnetars can be better understood if one considers the magnetostatic equations
as 
\begin{equation} 
{\nabla}.\vec{B}=0
\label{1}
\end{equation}
\begin{equation}
{\nabla}{\times}\vec{H}-\frac{1}{c}\vec{J}=0
\label{2}
\end{equation}
\begin{equation}
{\nabla}{\pi}= \frac{1}{c}\vec{J}{\times}\vec{B}
\label{3}
\end{equation}
where we have considered the absence of external forces and the constitutive equations and the total pressure are given respectively by
\begin{equation} 
\vec{B}={\mu}\vec{H}
\label{4}
\end{equation}
\begin{equation}
\pi={\rho}^{2}\frac{\partial}{{\partial}{\rho}}e
\label{5}
\end{equation}
\begin{equation}
P=\pi+\frac{1}{2{\mu}}B^{2}
\label{6}
\end{equation}
In the next section we shall examine the magnetostatic system above to the Serret-Frenet frame to investigate vortex magnetic filaments in magnetars.
\section{Magnetars and vortex filaments: the effects of torsion}
Let us now consider the magnetostatic equations in the Serret-Frenet (SF) frame $(\vec{t},\vec{N},\vec{b})$ where the magnetic field $\vec{B}$ can be written in terms of the SF frame along the unit vector $\vec{t}$ as $\vec{B}=B\vec{t}$. Here R is the radius of curvature and the curvature ${\kappa}(s)=\frac{1}{R}$ is constant where s is the measure of lenght along the vortex filament. This reasoning leads to
\begin{equation}
(\vec{B}.{\nabla})\vec{B}=B\frac{\partial}{{\partial}s}(B\vec{t})
\label{7}
\end{equation}
where the SF formulas are 
\begin{equation} 
\frac{\partial}{{\partial}s}\vec{t}={\kappa}(s)\vec{N}
\label{8}
\end{equation}
\begin{equation}
\frac{\partial}{{\partial}s}\vec{N}=-{\kappa}(s)\vec{t}+{\tau}(s)\vec{b}
\label{9}
\end{equation}
\begin{equation}
\frac{\partial}{{\partial}s}\vec{b}=-{\tau}(s)\vec{N}
\label{10}
\end{equation}
where ${\tau}$ represents the torsion of the filament. By considering the expression ${\nabla}{\pi}=0$ we obtain the following magnetostatic equations in the SF frame
\begin{equation} 
{\nabla}P=\vec{t}\frac{\partial}{{\partial}s}(\frac{1}{2{\mu}}{B}^{2})+\vec{N}(\frac{B^{2}}{R{\mu}})
\label{11}
\end{equation}
\begin{equation}
{\nabla}\pi=[\frac{B^{2}}{R{\mu}}-\frac{\partial}{{\partial}N}(\frac{1}{2{\mu}}{B}^{2})]\vec{N}-\frac{\partial}{{\partial}b}(\frac{1}{2{\mu}}{B}^{2})\vec{b}
\label{12}
\end{equation}
The MHD astrophysical application we consider here is a magnetic star in equilibrium under the action of it is own gravitational field. Therefore here we consider the external force of gravity
\begin{equation}
\vec{f}=-{\nabla}{\phi}
\label{13}
\end{equation}
and the magnetostatic equations become
\begin{equation}
-{\nabla}{\phi}= \frac{1}{\rho}{\nabla}{\pi} -\frac{1}{{\rho}c}\vec{J}{\times}\vec{B}
\label{14}
\end{equation}
\begin{equation}
{\nabla}{p}= \frac{1}{c}\vec{J}{\times}\vec{B}
\label{15}
\end{equation}
where
\begin{equation}
p:= {\pi}+{\rho}{\phi} 
\label{16}
\end{equation}
\begin{equation}
\frac{\partial}{{\partial}s}\vec{b}=-{\tau}(s)\vec{N}
\label{17}
\end{equation}
where ${\tau}$ represents the torsion of the filament. By substitution of equations (\ref{11}) and (\ref{12}) into the magnetostatic equilibrium equation we obtain
\begin{equation} 
\frac{J_{N}B}{{\rho}c}=\frac{\partial}{{\partial}b}(\frac{1}{2{\mu}}{B}^{2})-\frac{\partial}{{\partial}b}{\phi}
\label{18}
\end{equation}
\begin{equation}
\frac{J_{b}B}{{\rho}c}=-\frac{1}{\rho}[\frac{B^{2}}{R{\mu}}-\frac{\partial}{{\partial}N}(\frac{1}{2{\mu}}{B}^{2})]-\frac{\partial}{{\partial}N}{\phi}
\label{19}
\end{equation}
where we have considered the gravitational gradient ${\nabla}{\phi}$ written in the SF frame as
\begin{equation}
{\nabla}{\phi}= (\vec{t}\frac{\partial}{{\partial}s}+\vec{N}\frac{\partial}{{\partial}N}+\vec{b}\frac{\partial}{{\partial}b}){\phi}
\label{20}
\end{equation}
which substitution in the magnetostatic gave rise to
\begin{equation}
{\nabla}{\phi}=-[{\alpha}\vec{b}+{\beta}\vec{N}]
\label{21}
\end{equation}
where ${\alpha}$ and ${\beta}$ are given respectively by
\begin{equation} 
\alpha=\frac{1}{{\rho}c}[J_{N}B-\frac{\partial}{{\partial}b}(\frac{1}{2{\mu}}{B}^{2})]
\label{22}
\end{equation}
\begin{equation}
\beta=\frac{1}{{\rho}c}[J_{b}B+\frac{B^{2}}{{\rho}R}-\frac{B^{2}}{R{\mu}}-\frac{\partial}{{\partial}N}(\frac{1}{2{\mu}}{B}^{2})]
\label{23}
\end{equation}
The introduction of the torsion of the filament can be done by derivation of this equation with respect to s
\begin{equation}
(\frac{\partial}{{\partial}s}{\alpha})\vec{b}+(\frac{\partial}{{\partial}s}\vec{b}){\alpha}+(\frac{\partial}{{\partial}s}{\beta})\vec{N}+(\frac{\partial}{{\partial}s}\vec{N}){\beta}=-{\nabla}\frac{\partial}{{\partial}s}{\phi} 
\label{24}
\end{equation}
By making use of the SF equations we obtain
\begin{equation}
[\frac{\partial}{{\partial}s}{\alpha}+{\tau}{\beta}]\vec{b}+[\frac{\partial}{{\partial}s}{\beta}-{\tau}{\alpha}]\vec{N}-{\kappa}{\beta}\vec{t}=-\vec{t}\frac{{\partial}^{2}}{{\partial}s^{2}}{\phi} 
\label{25}
\end{equation}
Thus we are left with two scalar equations for ${\alpha}$ and ${\beta}$ and ${\phi}$ in terms of the torsion $\tau$
\begin{equation}
\frac{\partial}{{\partial}s}{\alpha}+{\tau}{\beta}=0 
\label{26}
\end{equation}
\begin{equation}
\frac{\partial}{{\partial}s}{\beta}-{\tau}{\alpha}=0 
\label{27}
\end{equation}
\begin{equation}
\frac{{\partial}^{2}}{{\partial}s^{2}}{\phi}-{\kappa}{\beta}=0 
\label{28}
\end{equation}
where to simplify matters we consider that the gravitational potential of the magnetars depend only upon the variable s. By considering regions inside the Sun for example where the torsion of vortex filaments are approximately constant we are able to reduce these two scalar equations to harmonic like equations 
\begin{equation}
\frac{{\partial}^{2}}{{\partial}s^{2}}{\beta}+{\tau}^{2}{\beta}=0 
\label{29}
\end{equation}
\begin{equation}
\frac{{\partial}^{2}}{{\partial}s^{2}}{\alpha}+{\tau}^{2}{\alpha}=0 
\label{30}
\end{equation}
A particular solution for these equations would be
\begin{equation}
{\alpha}={\beta}=e^{i\frac{\tau}{2}s}
\label{31}
\end{equation}
which written back in terms of the magnetic field implies the following relations
\begin{equation}
\frac{\partial}{{\partial}b}(\frac{1}{2{\mu}}{B}^{2})=\frac{\partial}{{\partial}N}(\frac{1}{2{\mu}}{B}^{2})=0
\label{32}
\end{equation}
and when R goes to infinity although still inside the star. Conditions (\ref{31}) implies that the magnetic field depends only on the parameter s. This particular solution presents a degeneracy in the current flows such that $J_{b}=J_{N}={\rho}e^{i\frac{\tau}{2}s}$ and the magnetic field becomes $B(s)= c e^{i\frac{\tau}{2}s}$ which presents clearly the contribution of torsion on the vortex filaments inside the magnetic star such as the sun. With these solutions for B and Js now the gravitational potential ${\phi}$ may be easily obtained by solving the equation (\ref{28}) or equivalentely
\begin{equation}
\frac{{\partial}^{2}}{{\partial}s^{2}}{\phi}-\frac{\kappa}{c}e^{i{\tau}s}=0 
\label{33}
\end{equation}
which yields the Hasimoto \cite{6,7} soliton transformation factor ${\psi}= {\kappa}e^{(i\int{{\tau}ds})}$ in the Schroedinger nonlinear equation for the constant torsion of the vortex filaments, on its first quadrature
\begin{equation}
\frac{{\partial}}{{\partial}s}{\phi}=-i\frac{\kappa}{c{\tau}}e^{i(\int{{\tau}ds})} 
\label{34}
\end{equation}
whose solution is 
\begin{equation}
{\phi}(s)= -
\frac{\kappa}{c{\tau}^{2}}e^{({i{\tau}s})} 
\label{35}
\end{equation}
This shows that in our particular solution torsion of the vortex filaments makes a major effect on the internal structure of the magnetic star.
\section{Conclusions}
Several phenomena in solar physics such as hot spots has made use of the twisted vortex filament structure. In this paper we generalize the investigation of the effects of torsion to vortex filaments to general types of magnetic stars which may include for example pulsars besides the sun itself. A more general solution of the dynamical equations of the vortex filaments obtained here maybe important for the understanding of internal structure of magnetars.


\begin{thebibliography}{7}
\bibitem{1} R. Ricca, Inflexional disequilibrium of magnetic flux-tubes ,Fluid Dynamics Research 36 (2005) 319. R. Ricca, Evolution and Inflexional stability of twisted magnetic flux tubes. Solar Physics 172 (1997) 241. R. Ricca, Phys. Rev. A (1999).
\bibitem{2} D.W. Moore and P.G. Saffman, The motion of a vortex filament with axial flow, Phil. Transactions of R. Soc. London 272 (1972)403.  
\bibitem{3} R. Ricca, The effect of torsion on the motion of a helical vortex filament, Journal of Fluid Mechanics, 237 (1994) 241.
\bibitem{4} J. Brat et al, Plasma Loops in the Solar Corona (1991) Cambridge University Press.
\bibitem{5} A. C. Eringen and G. A.  Maugin, Electrodynamics of continua II-Fluids and Complex Media (1990) Springer-Verlag.
\bibitem{6} H. Hasimoto, A soliton on a vortex filament, J. Fluid Mechanics 51 (1972) 477.  
\bibitem{7} W.K. Schief, Physics of Plasmas 10,7 (2003) 2677. W.K. Schief,J. Plasma Physics (2003) 65,6,465.
\end{thebibliography}
\end{document}